# Discovery of the neutron stars merger GW170817/GRB170817a and Binary Stellar Evolution

**Vladimir Lipunov[a,b], Victor Kornilov[a,b], Evgeny Gorbovskoy[a,b], Galina Lipunova[b], Daniil Vlasenko[a,b], I.Panchenko[b] Tyurina, N.V.[b], Grinshpun, V.[b]**

[a] Lomonosov Moscow State University, Physics Department, 119991, Vorobievy hills, 1, Moscow, Russia
[b] Lomonosov Moscow State University, SAI, 119234, Universitetsky pr., 13, Moscow, Russia

## ABSTRACT

*The Multimessenger discovery of the merger of two neutron stars on August 17, 2017, GW170817 / GRB170817a, accompanied by a gamma-ray burst and an optical kilonova, is a triumph of the ideas about the evolution of the baryon component in the Universe. Despite the current uniqueness of this observation, the obtained variety of experimental data makes it possible right now to draw important theoretical conclusions about the origin of the double neutron star, their merger, and the subsequent flare-up of the electromagnetic radiation. We present that the discovery of the merger at a distance of 40 Mpc is in full agreement with the very first calculations of the Scenario Machine (Lipunov et al. 1987). In modern terms, the predicted rate is ~ 10 000 $Gpc^{-3}$.*

## 1. Introduction

Neutron stars are the first class of the astronomical objects whose existence was predicted theoretically (Landau, 1932, Baade and Zwicky (1934) linked the formation of these objects with a supernova (SN) explosion and directly pointed to the supernova remnant of 1054 in the Crab nebula. In 1966, Zeldovich and Novikov found a physical process that could make these objects, with a radius of about 10 km, - microscopic in scale compared to the ordinary stars, - seen as bright sources of the electromagnetic radiation. This mechanism - the accretion of the surrounding gas onto a neutron star - was proposed by Shklovsky (1967) as an explanation of the nature of the brightest X-ray source Sco X-1. Almost at the same time, Kardashev (1964) and Pacini (1967) found another source of energy of the magnetized neutron star: its rotational energy stored during the collapse. Thus, the neutron stars, born at the tip of the pen, became a scientific hypothesis, directly confirmed with the discovery of radio pulsars (Hewish et al., 1968) and accreting X-ray pulsars (Schreier 1972, Giacconi et al. 1972).

After the discovery of the double radio pulsar (Hulse and Taylor 1974), it became clear that the collision of neutron stars was occurring in the Universe because the time of coalescence of the components of the binary was less than the Hubble time (Brumberg et al. 1975).

On August 17, 2017, 12: 41: 06.47 UT von Kienlin et al 2017 reported that the Gamma Burst Monitor (GBM) installed at the Fermi Observatory registered a short (2 s) gamma pulse that occurred 2 seconds after recording the gravitational-wave burst (GWB/GW) (Connaughton et al., 2017ab). Savchenko et al. 2017 a posteriori found a short and relatively weak transient at T0 with the S/N ratio larger than 3, coincident with the GBM trigger (Connaughton et al., 2017a).

Potential host galaxies list at preliminary 37+-12Mpc distance with brightness, face-on markers, SFR values was publishied by Cook et al. 2017, comparing with previous list of Dalya et al. 2017

The next day, 18 Aug 2017, the optical 1-m Swope telescope at Las Campanas Observatory was the first to report a new source SSS17a, 5.3" east and 8.8" north of NGC 4993 galaxy at a distance of ~ 40 Mpc (Coulter et al., 2017) - the third galaxy in Cook et al 2017 list.

It was independently detected by MASTER-OAFA auto-detection system as MASTER OTJ130948.10-232253.3 (Abbott et al.2017, Lipunov et al. 2017g) and confirmed by multicolor photometry that was obtained by several telescopes (Yang et al.2017, Melandri et al. 2017, Nicholl et al. 2017a, Tanvir and Levan 2017, Lipunov et al 2017cd, Chambers et al. 2017, Yoshida et al. 2017).

It is remarkable that the optical object, found 12 hours after the merger in NGC 4993, was not similar in behavior, brightness, or spectrum to any of the supernovae studied in the past. The soon obtained optical spectra (Drout et al., 2017, Lyman et al., 2017, Shara et al., 2017, see also Abbot et al. 2017) confirmed that the shell of the kilonova (KN) is dispersed at a speed of 100 000 km/s, which corresponds to the escape velocity on the surface of a neutron star.

Thus, on August 17, 2017, astronomers and physicists virtually simultaneously observed for the first time the collision of two neutron stars and its consequences in the NGC 4993 galaxy, not only in the gravitational-wave channel, but also in several ranges of electromagnetic radiation: from the gamma, X-ray, ultraviolet, to the optical and infrared ranges (Abbott et al. 2017).

Despite the uniqueness of this event, the variety of the experimental data makes it possible right now to draw important theoretical conclusions about the origin of the double neutron stars, their merger and the accompanying bursts of the electromagnetic radiation.

Below we will try to understand how this phenomenon corresponds to our ideas and what new knowledge is brought by this discovery.

## 2. Mergingology of the M-reaction.

The merger of the relativistic stars, neutron stars (NS) and black holes (BH), is the result of the evolution of massive binary systems (Tutukov and Yungelson 1972; van den Heuvel and Heise1972; Flannery and van den Heuvel 1975; Clark et al. 1979).

Blinnikov et al. (1984) first suggested that the destruction of a neutron star in a binary system can lead to a powerful electromagnetic flare similar to a gamma-ray burst (GRB). Lipunov 1993,2005 noted that the power of the energy released in the process of merger of the relativistic stars approaches the universal upper limit of the luminosity:

$$L_{max} = Mc^2/R_g/c = E_{Pl}/t_{Pl} = C^5/G \sim 2.4 \cdot 10^{59} \text{ erg/s} \tag{1}$$

and, hence, such events can be the most powerful explosions in the universe. Here $E_{Pl}$ and $t_{Pl}$ are the so-called Planck energy and Planck time. We emphasize that the limit (1) holds in the future theory of the quantum gravity, since the Planck constant falls out of the formula.

The variety of different types of mergers makes one speak of the science of "mergingology" (Lipunov 1998a). Depending on the masses of the stars involved in the `M-reaction', the following options are possible:

**NS + NS => GWB + sGRB + KN + NS**   if   $M_1 + M_2 < M_{ov}$ $\qquad(2)$

**NS + NS => GWB + sGRB + KN + BH**   if   $M_1 + M_2 \geq M_{ov}$ $\qquad(3)$

**NS + BH => GWB + sGRB + KN + BH** $\qquad(4)$

**BH + BH => GWB + BH** $\qquad(5)$

where $M_{ov}$ is the Oppenheimer-Volkoff limit which is not known exactly unknown now. Usually it is assumed that $M_{ov} \sim 2\text{-}3\, M_\odot \sim M_1 + M_2$. Consequently, the answer to the question, what products arise in the process of merger of the neutron stars, can be given by the observations of the mergers. Abbreviations GWB

There is one more type of the merger, which is potentially capable of producing a powerful gravitational-wave signal; it involves heavy O-Ne-Mg white dwarfs (Lipunov 2017a):

**WD + WD => GW + SNIa + NS**   if   $M_{OV} > 2M_{Ch}$ $\qquad(6)$

**WD + WD => GW + SNIa + BH**   if   $M_{OV} < 2M_{Ch}$ $\qquad(7)$

Now, thanks to the discovery of the colliding black holes (Abbot et al., 2016abcd) and, especially, the colliding neutron stars (Abbot et al. 2017), we can be confident that the science of the Relativistic Mergingology has become a part of the observational astronomy.

As was shown earlier by the method of the population synthesis of the binary systems, the merging black holes, although they merge less often, are detected more frequently (Lipunov et al. 1997a, b, c; Tutukov and Yungelson, 1993; see details in Lipunov et al. 2017).

The discovery of the colliding neutron stars and black holes (Abbot 2017), which occurred practically simultaneously and with interferometers of close sensitivity, is the strongest argument in favor of the fact that the binary systems that generated these phenomena have a common nature. These are the ordinary though massive stars capable of producing the relativistic remnants.

In the next sections, we will consider the astrophysical consequences to the recent observations. We present one of the possible scenarios for the evolution of a system of two massive stars leading to the merger of the neutron stars registered on August 17, 2017 by the gravitational-wave detectors of the LIGO/Virgo collaboration. We also discuss the physical nature of the objects possibly produced in the course of the mergers.

## 3. NS's Merging Rate.

Clark et al.,1979 first attempted to estimate the rate of merger of neutron stars in the Galaxy based on general ideas about the evolution of the binary systems up to the formation of relativistic stars in them. The estimate was rather approximate: $10^{-4}$ - $10^{-6}$ mergers per year. Why? Because the Merging Rate is the product of a large number of hard-to-estimate probability coefficients, like the Drake formula for the number of habitable planets in the Galaxy. Namely, Merging Rate

$$\text{MRate} = f \times \alpha \times \beta \times \delta \times \gamma \tag{8}$$

where $f$ is the Star Formation Rate in the Galaxy (the Salpeter Function),
$\alpha$ is the part of binary stars that can produce relativistic stars,
$\beta$ is the part of the stars that survived after the first SN explosion (it strongly depends on the anisotropy of the collapse or the so called kick-velocity),
$\gamma$ is the part of the neutron stars after the second SN explosion,
and $\delta$ is the part of the double relativistic stars, that can merge in the Hubble time.
Whereas the Salpeter function is more or less determined (with a factor of 2):

$$f(M) \approx 1 \, (M/M_\odot)^{-2.35} \, yr^{-1} \tag{9}$$

the remaining coefficients are known very approximately.

In the early 1980s, Kornilov and Lipunov,1983a,b proposed and implemented the population synthesis of binary stars by the Monte-Carlo method - the Scenario Machine. The main idea of the Scenario Machine was not only to use the various scenarios of the evolution of binary stars to predict the numbers and probabilities of unobservable processes, but, first, to try to achieve the selection of such evolutionary parameters that explain the observed stages of the binary systems in the most optimal way. Among such parameters, we can distinguish the speed of the kick during a SN explosion $V_{kick}$, $\alpha_{CE}$, the efficiency parameter of the common shell (van den Huevel, 1996), and $\alpha_q$, the parameter of the distribution function of binary systems on mass ratio $q = M_2 / M_1 < 1$:

$$\varphi(q) \sim q^{\alpha_q} \tag{10}$$

The first calculation of the merger frequency using the population synthesis (Scenario Machine) was carried out in 1987 by the Monte-Carlo method (Lipunov et al, 1987, hereafter LPP87). It was found that the mergers of neutron stars in a galaxy of our type (that is, in a galaxy with the star formation rate of one solar-type star per year according to the Salpeter function) occur once a year in a sphere of 20 Mpc (LPP87, see the caption to figure 1, case "e"). Since there are ~ 10000 galaxies of our type in the volume of this radius surrounding our Galaxy, the average frequency of mergers per galaxy of our type is $10^{-4}$ yr$^{-1}$.

A similar estimate by analytic methods was later obtained by Hills et al.,1990 and Tutukov and Yungelson 1993. The latest attempt to get the merger rate by simple analytic evaluation was undertaken by Bethe and Brown, 1999. It should be noted that attempts to obtain the merger rate directly on the basis of observations of radio pulsars led to a prediction of the merger rate 2 orders of magnitude below: ~ $10^{-6}$ yr$^{-1}$ (Phinney, 1991; Narayan et al. 1991). This discrepancy gave rise to a certain skepticism of the investigators far from the population synthesis field in relation to the possibility of obtaining more or less reasonable estimates of this process.

Meanwhile, the subsequent calculations of the Scenario Machine conducted in the 1990s (Lipunov et al. 1995, 1997) confirmed the first estimate of the merger rate. On the other hand, the so-called observational estimates obtained from observations of binary radio pulsars gradually grew and approached the result of the populational synthesis (Curran and Lorimer 1995; Van den Heuvel and Lorimer 1996; Bailes 1996; Burgay et al. 2003) (Fig. 1).

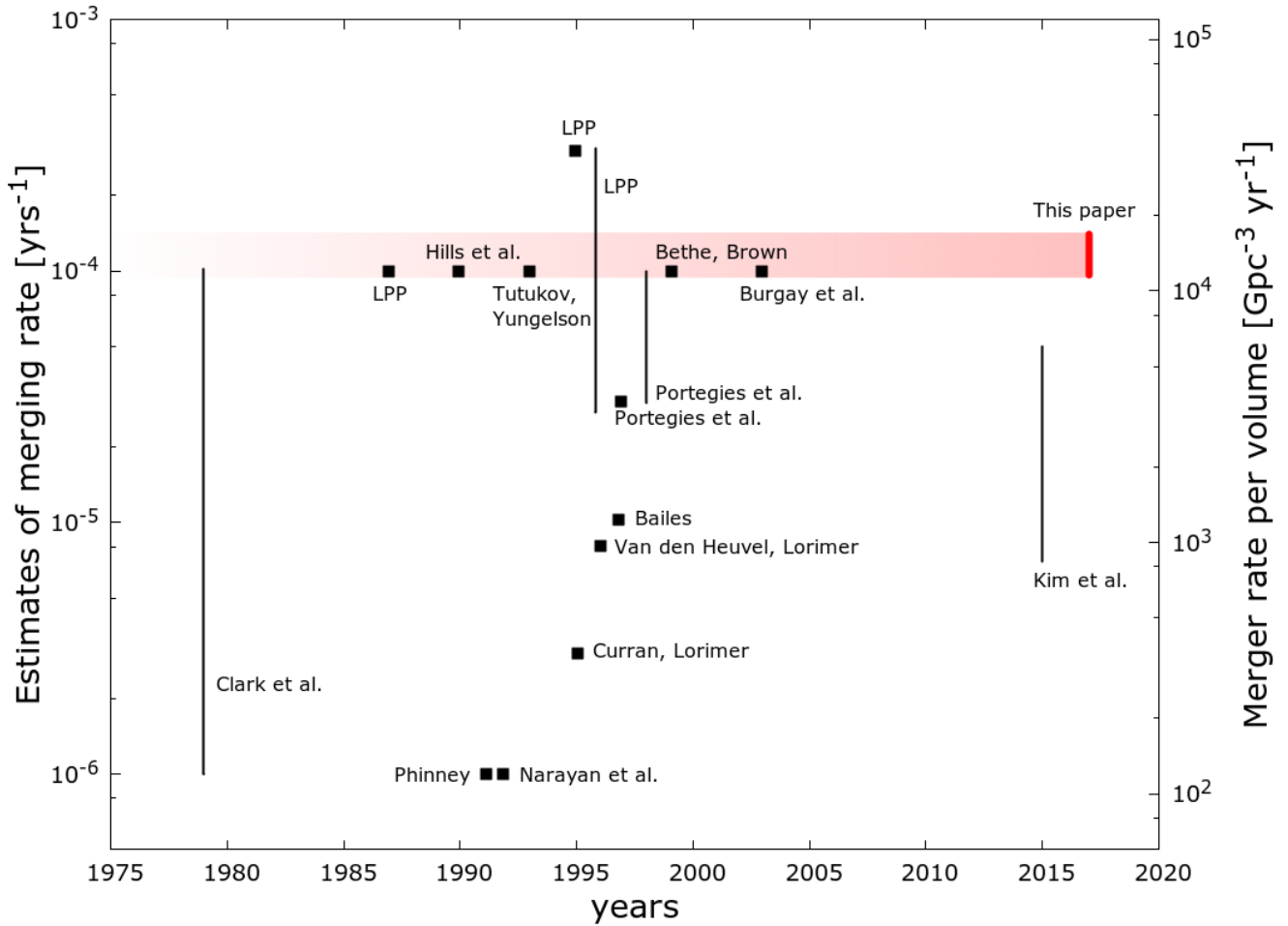

Fig.1. *History of estimates of the neutron stars merger rate. From left to right: Clark et al. 1979; LPP - Lipunov et al., 1987; Hills et al. 1990; Phinney 1990; Narayan et al. 1991; Tutukov and Yungelson 1993; Lipunov et al. 1994; Lipunov et al. 1995a; Lipunov et al. 1995b; Curran, Lorimer, 1995; Van den Heuvel, Lorimer 1996; Bailes 1996; Portegies Zwart and Spreeuw 1996; Portegies Zwart and Yungelson 1998; Bethe and Brown 1999; Burgay et al. 2003; Kim et al 2013.*

At the same time, several groups periodically published various estimates that were tending toward a decrease in the merger rate in the volume of sensitivity of the gravitational-wave antennas under construction (Kalogera and Lorimer, 2000, Belczynski et al., 2002).

Does not the spread of 1.5 to 2 orders of magnitude in different studies on the mergers' rate by different groups show the true accuracy of such calculations? It would, but only for those calculations that do not make sure that the objects, observed from the Earth, do not disappear in their artificial galaxy.

The question arises, why, year after year, in spite of progress in our understanding of the evolution of the binary stars, the Scenario Machine gives the same answer to the question of how often neutron stars merge? In one of the works (Lipunov 2005), it is explicitly stated that the merger rate in a galaxy of our type cannot be less than $10^{-4.5}$ yr$^{-1}$, otherwise the binary un-recycled radio pulsars would disappear from the sky.

As it was repeatedly stressed by one of the authors (VML), the one of the most reliable observables in this science is the dimensionless relative number of observed un-recycled radio pulsars in pairs with a neutron star, normalized to the total observed number of radio pulsars (single and double).

There are two reasons for this:
(1) These systems are the direct ancestors of the merging neutron stars.
(2) Their observed number, affected by the selection effects (such poorly known parameters of pulsars as the average distribution of kick directions, the spin-down law, the death line, and the spectral characteristics), does not depend on whether the pulsar is isolated or it is in a binary system. Of course, we refer not to recycled millisecond pulsars, which have a different life and destiny.

Lipunov et al. 1997c demonstrated, how the relative number of neutron stars in binary systems with radio pulsars depends on the kick velocity during the SN explosion (see their Fig.1).

The dependence is particularly sharp for the neutron stars with radio pulsars - potential progenitors of the events at the gravitational wave antennas. This is due to the fact that such systems have managed to survive in two SN explosions, that is, the recoil momentum has acted twice.

The merger rate versus the kick velocity was calculated in the same work. We have combined the two graphs to get the dependence of the merger rate on the relative number of radio pulsars with neutron stars (Fig. 2). The vertical line corresponds to the modern statistics of the radio pulsars from the constantly updated catalog (http://www.atnf.csiro.au/research/pulsar/psrcat/ by Manchester et al. ,2005).

We excluded the millisecond pulsars, the so-called recycled pulsars, from the number of binary pulsars: they have a different history of evolution of the period and magnetic field comparing to the single pulsars.

Consequently, there are 4 binary pulsars with neutron stars the at the moment, so their relative number to single pulsars (PSR+NS)/PSR is equal to 0.16%. We use following pulsars PSR J0737-3039B, PSR J1906+0746, PSR 1755-2550, PSR J1930-1852 (see Tauris et al., 2017)**.**

PSR J1930-1852 lies on the board of "young" pulsars. At this time the spin-up , that pulsar could be affected in double system, was short-term and didn't led to change the character of his evolution (Swiggum et al., 2015).

There are not all non recycled binary pulsars, but the most evident.

As result we have next estimation NSs Merger Rate: (1-1.5) $10^{-4}$ yr$^{-1}$ for **SFR (M > 1M$_\odot$) = 1** (M/ M$_\odot^{1.35}$ ) per year for Maxwell and Lyne-Lorimer kick velocity distribution(Fig.2).

This is equivalents **12000 - 16 000 yr$^{-1}$ Gpc-3** in terms Merging Rates per local **10$^9$** pc volume.

Last estimation is in good agreement with distance of the first NSs merging distance detected 17 Aug 2017 ~ 40 Mpc during 1/3 yr LVC observations (set O2, see Abbot et al., 2017, LVC Phys.Rev).:

**LVC Merging Rate  ~  3/4π/3 · (40Mpc)$^3$  ~ 12 000 Gpc$^{-3}$**      (11)

We not include important selection effect connected with the difficulties of finding wide separated binary pulsars (see Manchester & Taylor, 1977). In this Scenario Machine the calculation merging time is not important!

From the fact that this is the lower estimate of the fraction of double pulsars with normal evolution we conclude that the rate of mergers cannot be much lower than $10^{-4}$ per Milky Way.

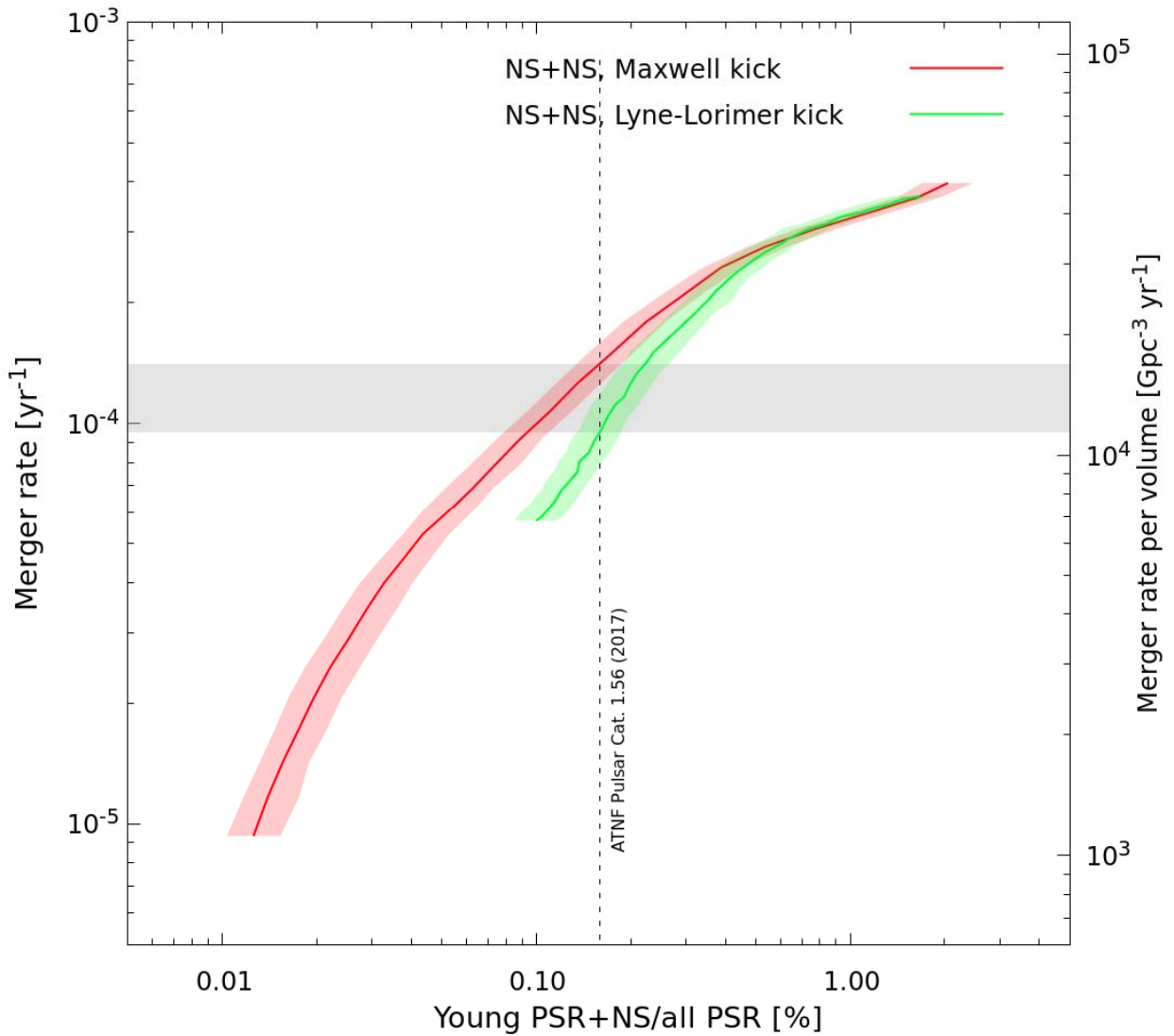

*Fig. 2. The merger rate of neutron stars in a galaxy with a constant Star Formation Rate of 1 M$_\odot$/yr (the Milky Way Galaxy Solpiter function) versus the relative number of radio pulsars with neutron stars for two distributions of the kick velocities (Maxwell and Lyne-Lorimer ). Bluring of the line caused by different initial mass ratio slope: 0<=a <=2 . The relation is constructed according to the calculations of the Scenario Machine (Lipunov et al., 1997c). The vertical line corresponds to the statistics of the pulsar catalog (Manchester, et al., 2005), taking into account only young double pulsars (Özel and Freire, 2016).*

The results of the Scenario Machine can be also presented as shown in Fig. 3. We plot the merger rate of the neutron stars as a function of the size of the sensitivity horizon (see figure 6 in Lipunov and Pruzhinskaya, 2014) and put the event GW170817. The distance to the galaxy NGC4993 is marked with a shaded region. As we see, the event of August 17, 2017 with a distance of 41 ± 5.8 Mpc is in good agreement with the predicted rate of several events per year ~ 2-4 yr$^{-1}$. We must remember that that total observation time of LIGO/VIRGO set O2 is about ~1/3 yr (Abbot et al., LVC, Phys.Rev. 2017).

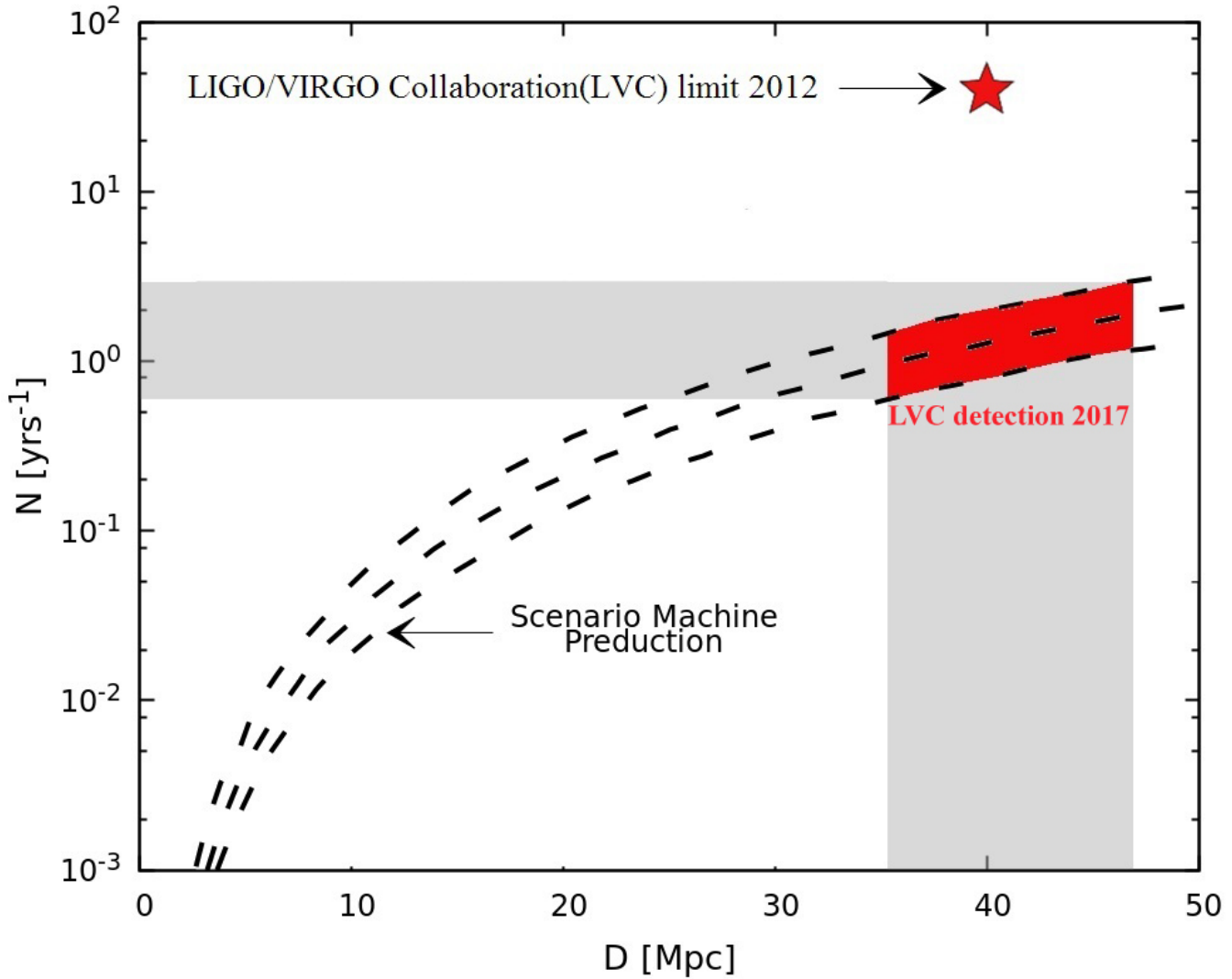

**Fig.3.** *The merger rate of the neutron stars as a function of the size of the sensitivity horizon (Lipunov and Pruzhinskaya, 2014). The red box is the GW 170817 NS's merging in NGC 4993 (Abbot et al., 2017). The distance to the event GW170817 is shown by the shaded region. The distance to NGC 4993 was measured using the Tully-Fisher method, $D_{TF} = 41\pm5.8$ Mpc . The asterisk shows the LIGO S6 limit (Abadie et al., 2012).*

## 4. Star formation and NS's merging in NGC 4993.

Apparently, astrophysicists have observed the collisions of neutron stars for about 40 years as short gamma-ray bursts. However, until now, it was, although the most probable, but still, only a hypothesis explaining their origin. After the discovery of the gravitational-wave pulse GW170817 by the LVC collaboration and the short gamma-ray pulse 2 seconds later (GRB170817A) by the Fermi observatory, we have a direct proof of this hypothesis. Moreover, after the discovery of the optical radiation in the bright nearby galaxy NGC4993, we have a direct possibility to understand the evolutionary status of the binary star system that generated the double neutron star, which on August 17 of 2017 gave birth to a gravitational wave phenomenon accompanied by an electromagnetic flare.

NGC 4993 is an early-type galaxy S0 or early E (Sadler at al. 2017). Such galaxies are characterized by a suppressed star formation (Karachentsev et al., 2013), which is confirmed in the case of NGC 4993 by direct ultraviolet and infrared spectral data. Thus, the ultraviolet observations suggest the rate of the formation of stars with a mass of more than 5 $M_\odot$ **SFR (NUV) ~ 0.003 $M_\odot$/yr** . This is consistent with the upper limit obtained from the infrared observations . It is obvious that massive stars are capable of generating neutron stars in this galaxy for already several billion years. This means that the registered gravitational-wave event cannot be associated with the collapse of the fast-rotating core of a massive star. In other words, we are dealing with a merger in an old system of two neutron stars several billion years old (Lipunov, LVC GCN 21621). We shall prove this statement by the method of population synthesis of the binary systems specifically for the galaxy NGC 4993.

The mass of the galaxy is estimated as $10^{10.64}$ $M_\odot$ (Ogando et al. 2008), which is characteristic of galaxies of the S0 type.

We need to make a few remarks about the calculations made with the Scenario Machine more than 20 years ago (Lipunov et al., 1995a,b; Lipunov and Pruzhinskaya 2014.). The following idealized situation was simulated. A galaxy with a mass of $10^{11}$ solar masses was formed simultaneously. At once, the star formation is completed, and the close binary stars (that is, the half of all stars) begin to evolve according to a chosen scenario. In fact, this is a fairly accurate model of an elliptical galaxy. In these galaxies, star formation dies quite quickly, since the interstellar gas, which feeds the star formation, is practically absent in these galaxies. Fig. 4 shows the resulting history of the merger rate. The two gaps on the curve are artifacts associated with the factitious division of binaries according to the Webbink classification (Webbink 1979, Lipunov et al.1987, Lipunov et al. 1996). Recall, that in binary systems, the filling of the Roche lobe occurs at different stages of nuclear fusion in the center of the stars, depending on the distance between the components and their masses. In systems of type A, the closest ones, the filling of the Roche lobe occurs at the time of burning of the hydrogen in the star center. In systems of type B and C, mass exchange occurs during the thermonuclear fusion of the hydrogen in a shell surrounding the core or during the stage of the red supergiant. Systems of type D are not close and do not undergo the mass exchange.

It is clear that there is no sharp boundary between the types in the nature and, in fact, such dips (see Fig.4) are absent. Keeping this in mind, the entire curve in Figure 4 (Lipunov et al.1987, Lipunov et al 1995), starting at 200 Myr, is described approximately by the law Rate ~ $t^{-1}$, which is a natural consequence of two facts (Lipunov et al. 2011). The first fact is that the initial distribution of binary systems on the semiaxes *a* is described by a flat logarithmic law (Popova et al. 1982, Abt 1983):

$$dN \sim \varphi(a)da \sim (1/a)\, da \qquad (12)$$

Since the double neutron star had two SN explosions in its history and perhaps two common envelopes, one could suggest that this law is violated for relativistic systems. But numerical calculations show that apparently this does not occur, and distribution (4) is not greatly distorted. The reason for this is simple; the law (4) implies only one thing: among the binary systems there is an approximate uniformity of diversity, meaning that the number of very close systems is roughly equal to the number of close and wide ones. Each "octave" of separation values contains the same number of systems. Therefore, if any systems become too close due to the SN explosions or common envelopes, there are always initially wider pairs that will take their place.

Now, let us recall that the change in the distance between the components due to the emission of the gravitational waves is determined by the Einstein formula (Landau and Lifshitz 1975)

$$da/dt \sim a^{-3}(M1 +M2)/M1M2 \sim a^{-3}, \qquad (13)$$

As the masses of all neutron stars are close, we forget about them and get $t \sim a^{-n+1}$. Accordingly, the merger rate is

$$MRate \sim \varphi(a)\, da/dt \sim t^{-1} \qquad (14)$$

It is remarkable that this law holds for any power law of the decreasing semiaxis on time (Lipunov et al., 2011). Using the results of Lipunov et al. (1995a) for estimates of the merger rate at times longer than 30-50 My, one can adopt the following analytic approximation:

$$MergingRate = 10^{-4}\, t^{-1}\, yr^{-1} / (10^{11}\, M_\odot) \qquad (15)$$

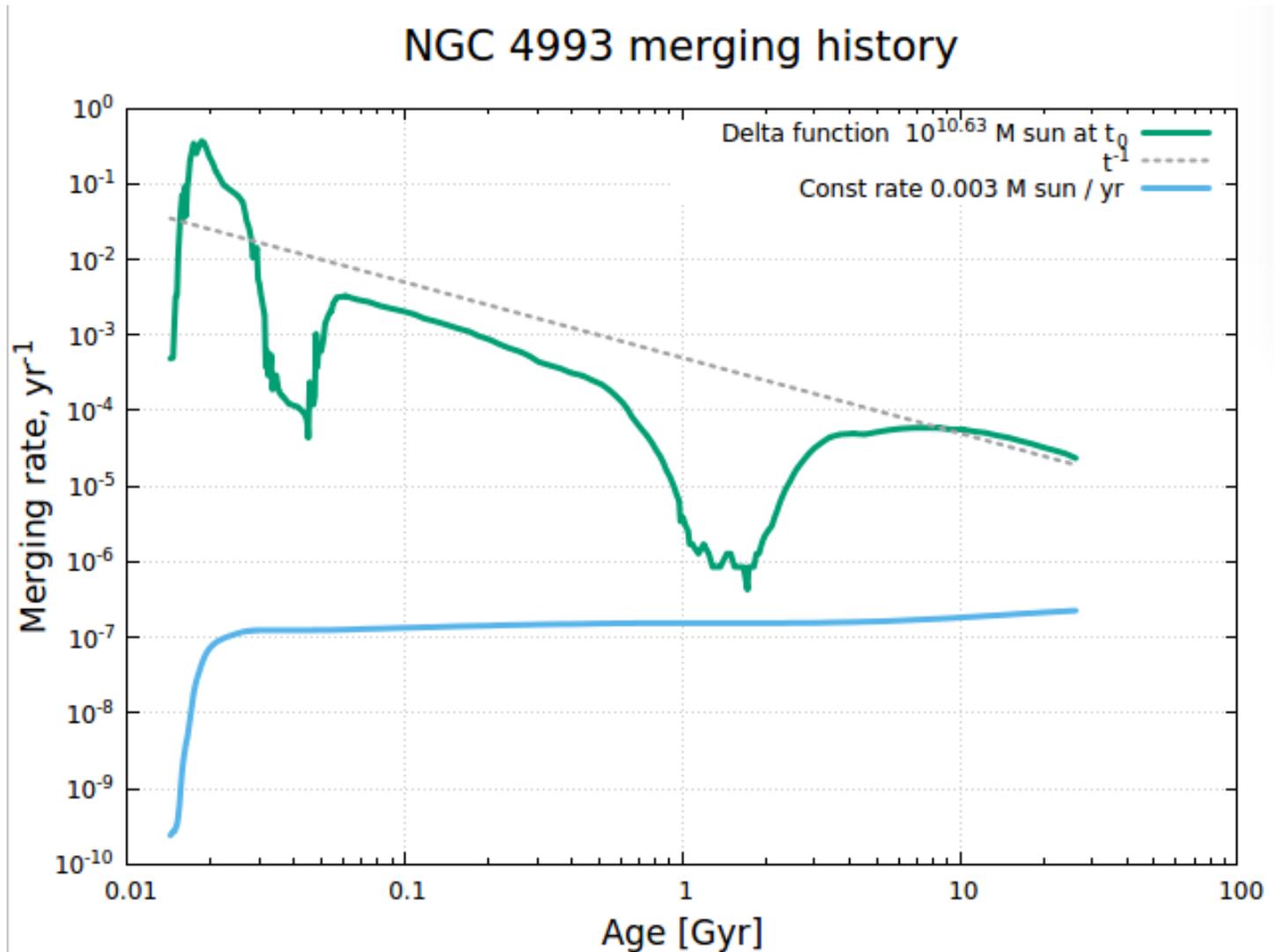

**Fig 4.** Evolution of the merger rate of neutron stars with the age of the galaxy, similar to NGC4993. It is assumed that the mass of the galaxy is $10^{10.63}\, M_\odot$, and the star formation rate is 0.003 $M_\odot$/yr. The green curve is plotted for an elliptical galaxy, with all stars formed at the same time (Lipunov et al., 1995a). The gray dotted line is the approximation by the law of reverse proportionality from the age.

As we noted, in respect of the history of star formation, the galaxies of the S0 type are closer to elliptical galaxies with a slight contribution of constant star formation. Fig. 4. presents a model of the history of the merger rate for such a galaxy with the star formation rate of **SFR ~ 0.003 M$_\odot$/yr** (from the NUV range, Sadler at al., 2017 LVC GCN 21645). It can be seen that at the present time the merger rate of neutron stars in the recently-born systems is two orders of magnitude lower than the contribution of binaries from the primary star-formation epoch. Thus, it is most likely that the ancestor binary system of the event GW170817 was formed about several to 10 billion years ago (Lipunov 2017 GCN LVC 21621).

In Fig. 5 we present one of the possible scenarios for the evolution of a system of two massive stars leading to the merger of the neutron stars registered on August 17, 2017 by the gravitational-wave detectors of the LIGO/Virgo collaboration. This evolutionary track is one of the many millions of those used in the population synthesis (Lipunov et al. 1996). The track is reproduced using the open online version of the Scenario Machine (Nazin et al. 1998).

| M1 M | | M2 M | A R₀ | T mln.yrs | Comments |
|---|---|---|---|---|---|
| 30.00 | | 15.00 | 290.00 | 0.00 | Two MS Stars |
| 28.17 | | 14.71 | 304.30 | 5.04 | Blue Giant + MS |
| 26.52 | | 14.68 | 316.70 | 5.54 | Blue Supergiant + MS |
| 11.69 | WR | 15.00 | 340.90 | 5.54 | WR + H Shell + MS |
| | SN II | | | 5.54 | SNII explosion |
| 1.40 | NS "E" | 15.00 | 809.90 | 5.72 | Fast NS Ejector + MS |
| 1.40 | NS "E" | 14.47 | 837.00 | 11.52 | Slow NS Ejector + MS |
| 1.40 | NS "SA" | 13.81 | 873.50 | 12.67 | NS Super Accretor + Red Giant |
| 1.40 | NS "A" | 13.15 | 347.30 | 12.67 | NS spin up during Common Envelope stage |
| 1.58 | NS "P"    WR | 4.43 | 2.79 | 12.68 | NS Propeller + WR |
| | NS "P"    SN Ib | | | 12.68 | Second SN explosion |
| 1.58 | NS "E"    NS "E'" | 1.40 | 23.54 | 13.04 | Two Ejecting Neutron Stars (Binary PSR) |
| 1.58 | NS "P"    NS "E'" | 1.40 | 17.50 | 628.20 | Propeller + Ejector (PSR) |
| 1.58 | NS "P"    NS "P" | 1.4 | 17.35 | 650 | Propeller+Propeller |
| | Coalescence | | | 650.00 | NS's Merging |
| | BH | 2.98 | | 10980.00 | Kerr Black Hole |

***Fig.5.*** *A possible evolutionary track of a binary star, leading to the merger of the neutron stars 10 billion years after the birth of the system. Here, in addition to the obvious designations, accepted also are the following: MS is a main sequence star, WR is a helium star of the Wolf-Rayet type, and E, P, A, SA are the ejectors, propellers, accretors, and super-accretors - the evolutionary stages of a magnetized neutron star (Lipunov, 1992).*

## 5. BH's and NS's realation merger: predictions and observations.

The first event of LIGO on September 14, 2015 was the merger of black holes. Qualitatively, the possibility of a priority of a black hole event was argued by Tutukov and Yungelson (1993).

The simulations with the Scenario Machine left no doubt that a merger of black holes would be the first to be discovered (Lipunov et al., 1997a, b, c). This has been shown for all possible scenarios with a wide variety of assumptions about the key parameters of the evolution.

Two years after the two observational semi-annual sets of the LIGO/VIRGO collaboration, there is no doubt about the fourfold preponderance of black holes on the detector, which is associated with a larger amplitude of the gravitational wave signal measured at the Earth.

The advantage of the event rate $ER_{BH+BH}/ER_{NS+NS} \sim 4$ is not as significant as predicted in Lipunov, 1997a and agrees even less with our recent estimates of the event rate (Lipunov et al. 2017a). However, one must bear in mind that this ratio depends strongly on the relative sensitivity of the LIGO/VIRGO detectors at low frequencies (~ 100-200 Hz) and high frequencies (~ 600-800 Hz).

If we assume that in the observations of O2 (spring-summer 2017) the sensitivity is increased more at higher frequencies than at low frequencies, then the ratio at the interferometer can become much smaller than the theoretically estimated one by several times. This is also not quite enough, but it is already close to what was predicted by the Scenario Machine in 1997a (so called `Dinosaur Head' ).

## 6. GW 170817 on the Gravitational Wave Sky.

Lipunov et al., 1995b, using the morphological catalog of Tully (1988) of the nearest galaxies and the results of the population synthesis of binary stars, plotted the probability distribution of mergers. It was assumed that star formation was long over in elliptical galaxies located at a distance <50 Mpc, and the merger of neutron stars was on its way in accordance with the Green function calculated by Lipunov et al.( 1995a ). In the spiral and irregular galaxies, the star formation rate was assumed to be constant and equal to the rate observed in our Galaxy,

 1 $M_\odot$ / yr / $10^{11}$ $M_\odot$, multiplied by the mass of a galaxy. The results of the calculations are reproduced in Fig. 7, together with the location of the first detected merger of neutron stars in the galaxy NGC4993.

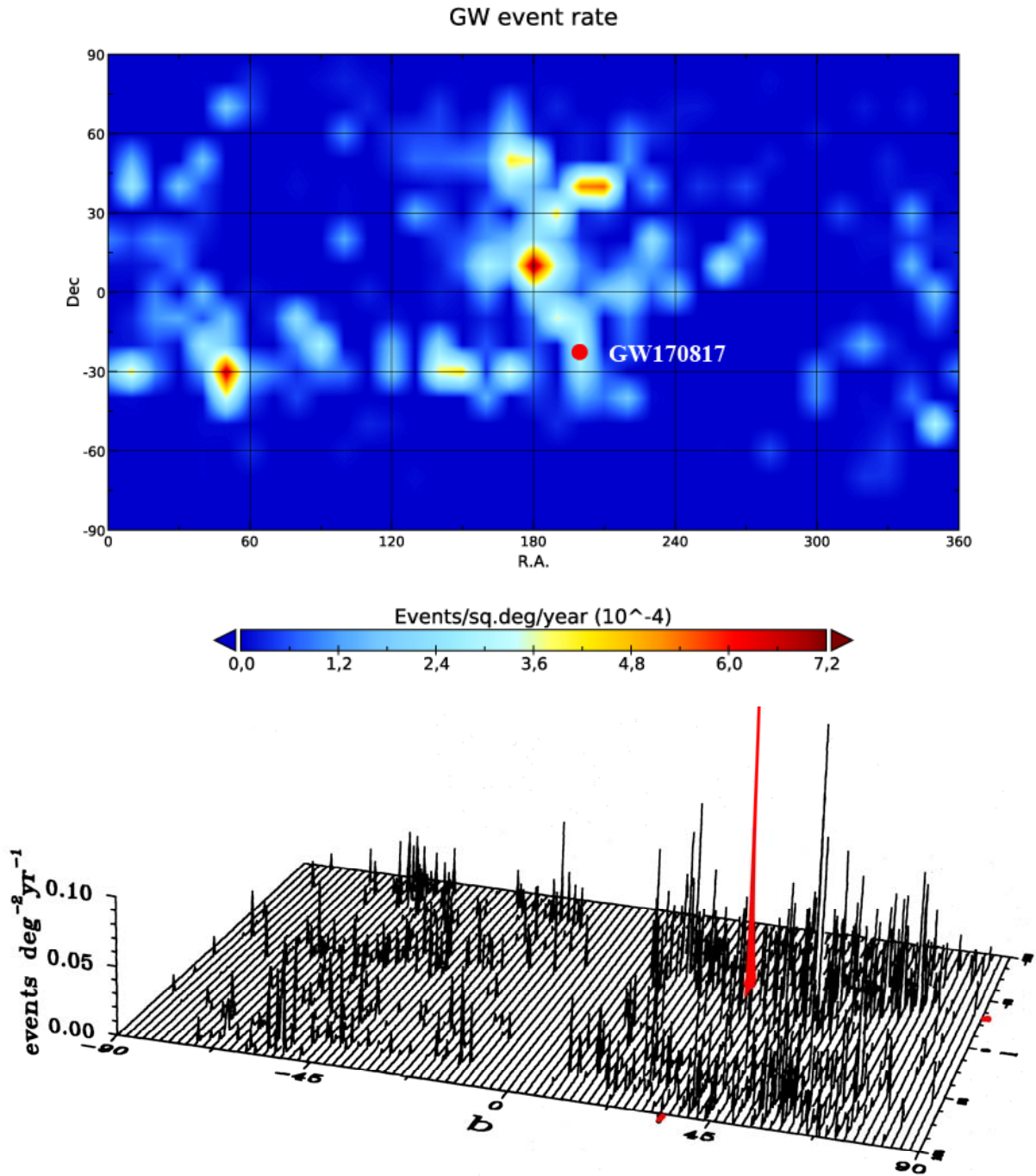

*Fig. 6.* Expected frequency of neutron star mergers across the sky in units of events per year per square degree *(Lipunov et al. 1995b)*. The total number of mergers is 3 events per year from a sphere of 50 Mpc. The Tully Catalog of Nearby Galaxies is used *(Tully 1988)*. The axes are the galactic coordinates. The red circle marks the event of August 17, 2017 in the galaxy NGC 4993, which falls into the `thicket' in the sky.

## 7. Kilonova as a standard candle.

During the merger, a part of the matter from neutron stars can be ejected and this may result in the radioactive decay of the synthesized heavy elements. In such a way a phenomenon of the so-called kilonova can be produced (Li and Paczynski 1998; Metzger et al., 2010; Tanvir et al., 2013; Berger et al 2013). Lipunov et al. (2017ApJletter) compare the observed optical luminosity of GW170817 kilonova with the optical luminosities of the other possible kilonovae observed earlier for GRB 130603B (Tanvir et al. 2013) and GRB 080503 (Perley et al. 2009).

**Table 1. Possible kilonovae brightness.**

| Name | $Z$ | $D_L$ (Mpc) | DM | Obs. | Band | Maximum visual magnitude, (mag, AB) | Absolute magnitude, Flat spectrum (mag, AB) | $F_{iso}$ (erg/s) | Ref |
|---|---|---|---|---|---|---|---|---|---|
| Kilonova NGC 4993 | 0.0098 | 42.5 | 33.14 | MASTER | W | 17.3 | -16.03 | $10^{42}$ | This paper |
| GRB 130603 | 0.3560 | 1911.9 | 41.41 | HST | H | 25.73 | -15.35 | $5.50 \times 10^{41}$ | Tanvir et al (2013) |
| GRB 080503 | 0.561* | 3290.5 | 42.59 | Gemini/Keck | r | 25.48 | -16.62 | $1.78 \times 10^{42}$ | Perley et al (2009) |

*We use the redshifts for the two galaxies that are visually closest to the GRB position (see discussion in the text).

The source spectra are unknown and, following Perley et al.,2009, we use a flat spectrum to estimate the k-correction. Hence, $M = m - DM + 2.5 \log(1+z)$, where DM is the distance modulus. Furthermore, for the flat spectrum sources the absolute magnitudes in all filters are equal.

We assume that the redshift of GRB 080503 is equal to the redshifts of the two galaxies visually closest to it. However, there is still no consensus regarding the host galaxy of GRB 080503. Perley et al. 2009 suggest that the nearest galaxies have no connection with GRB 080503, due to the large angular separations. However, we believe that any of these galaxies could be the GRB's host because a NS+NS system acquires a huge kick velocity during two SN explosions, allowing it to escape from a galaxy before the merger. In any case, this is the unique and best possible estimate of the redshift for the GRB.

The closeness of the observed absolute magnitudes and characteristic luminosities of three kionova (Table 1) evoke the old idea about regarding gamma-ray bursts as the standard candles (Lipunov et al. 2001). However, the kilonovae in question accompany the short gamma-ray burst events. The situation is rather similar to that of the Type Ia supernovae. Both kinds of events may represent collisions of the compact stars: binary white dwarfs and binary neutron stars in the case of supernovas and kilonovas, respectively. In addition, the mass of a collapsing object is crucial in the both processes: there is the Chandrasekhar limit for SN Ia and the Oppenheimer-Volkoff limit for kilonovae. This reasoning may be too naïve given that the densities of objects differ by a millionfold. We also note that a kilonova is much fainter than a SNIa. The kilonovae do not have simple spectral lines and can be outside of galaxies, which is expected because the binary neutron stars can escape from galaxies due to the kick velocity. In this case, the brightness of a kilonova is the only indicator of its distance!

To conclude, the optical radiation plays an important role in the energy balance of SN Ia, whereas this is by no means evident in the case of kilonovae, for which the total energy release in the IR and other ranges of the spectrum can be more important.

## 8. Black Hole or Magnetar?

As we see, in the case of the merger of August 17, 2017, we are dealing with the very old neutron stars that have lived for ~ $10^{10}$ yrs. We know from observations of millisecond pulsars, whose age is also estimated as billions of years (Lipunov 1992), that all of them have substantially weaker magnetic fields than the active pulsars. Indeed, the estimates on the spin-down of the rotation give $10^8$-$10^{10}$ G. Consequently, we cannot expect a pronounced electromagnetic precursor caused by the rapid orbital rotation (Lipunov and Panchenko 1996). We note that this is the case for the mergers in all elliptical galaxies. The different situation is in the spiral and irregular galaxies. There, about 40 per cent of the merging neutron stars have a comparatively small age of ~ 4 $10^7$ yrs, and the appearance of a precursor in the radio, optical, and X-ray range is quite expected.

After the beginning of the merger and the formation of the differentially rotating object, strong magnetic fields can be generated, up to the values characteristic for the magnetars (Usov 1992). What happens to the neutron star after the merger? A rapidly rotating body with a mass of ~ 2.5 - 3.5 $M_\odot$ is formed. Regardless of the actual Oppenheimer-Volkoff limit for a non-rotating neutron star, the rotating object maintains for some time its equilibrium by the centrifugal forces and also by the pressure of the neutron liquid. This type of object, a spinar, was first considered in the 1960s in connection with the problem of the existence of supermassive stars. We write the M-reaction for such merger as:

**NS + NS => Spinar => NS or BH**                                              (15)

Later, the spinars were discussed in the context of the collapse of the stellar-mass objects (Lipunov 1985; Lipunova 1997). The lifetime of such an object is determined by the dissipation time of the angular momentum. Lipunov and Gorbovskoy 2007, 2008 constructed a model for the evolution of a spinar taking into account all the relativistic effects associated with the collapse: the entrainment of the reference frames, the disappearance of the magnetic field during the collapse into a black hole, and the time dilation. It was shown that plateaus on the light curves of GRBs can be explained by a formation of a transient rotating compact object.

The dissipation of the rotational energy is caused by the magnetic viscosity or the magnetic braking. It turns out that the centrifugal barrier works for times from several fractions of a second to thousands of seconds depending on the magnitude of the spinar's magnetic field. This explains why the gamma-ray bursts last thousands of times longer than the free-fall time at the gravitational radius or the radius of the neutron star: $t$ ~ 10 km /100000 km/s ~ $10^{-4}$ s. In reality, the typical duration of a short GRB is of the order of $\Delta t$ ~ 1 s, which leads to the energy release in the model of the coalescing neutron stars of the order of 0.1 $Mc^2 / \Delta t$ ~ $10^{53}$ erg / sec. However, the brightness of short GRBs is 2-3 orders of magnitude lower, which can be explained by the small efficiency of the conversion of the gravitational energy of the collapse into the gamma radiation.

Fig. 7 shows the evolution of the energy release of the spinar along with the observational data (see below). The dimensionless Kerr parameter of the spinar is set to 1.5. The curve with the smooth decay at the end corresponds to the case when a neutron star is formed, the curve with the spike at the end demonstrates the evolution to a black-hole (different parameters in the equation of state are adopted in the two models, and the $M_{ov}$ limits differ, see Lipunov and Gorbovskoy 2007). During the characteristic time of the magneto-dipole evolution, which we will call the plateau time $t_{plato}$, the spinar looses the energy according to

$$L_{md} = \kappa_t \frac{\mu^2}{R_t^3} \qquad (16)$$

where $\kappa_t \sim 1$ and $R_t$ is the characteristic radius of interaction between the magnetic field and the ambient plasma. In a case of very dense ambient plasma, the characteristic magnetic torque radius equals the corotation radius: $R_t = R_{cor}$, and, in the opposite case, it equals the light-cylinder radius $R_{light}$ (Lipunov 1987, 1992).

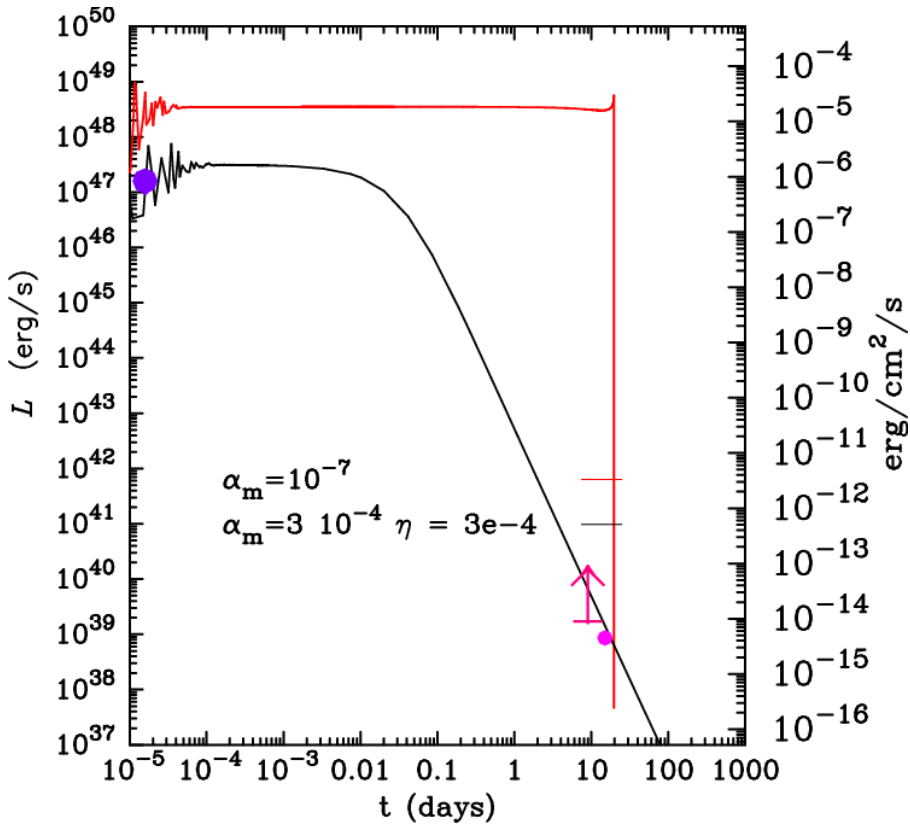

*Fig.7.* *The magneto-dipole losses of of the spinar calculated following the equations proposed by Lipunov and Gorbovskoy (2008) for the characteristic magnetic torque radius equal to the corotation radius: $R_t=R_{cor}$. The dots and the lower limit are the values reported (Troja et al. 2017). The black curve shows evolution of the spinar to a neuton star: the power of the spinar energy release is multiplied by the efficiency factor $\eta = 3 \cdot 10^{-4}$. The red curve with a spike at the end shows the evolution of the spinar to a black hole. The right axis shows the observed flux assuming the distance to the merger 40 Mpc.*

The characteristic values of the luminosity during the plateau and the duration of the plateau have different views depending on the choice of the torque radius $R_t$. If the medium around the spinar is so dense that the accretion proceeds on to it, $R_t = R_{cor}$ and:

$$L_{md} = \frac{2\alpha_m \kappa_t k^5 x_{ES}^{12}}{a^5} \frac{c^5}{G} = 1.8 \times 10^{46} \, \text{erg/s} \times \frac{\alpha_{-10} \kappa_t k^5 (x_{ES}/0.5)^{12}}{a^5} \qquad (17)$$

$$t_{plato} = \frac{1}{2} \frac{Rg}{c} \frac{a^3}{\alpha_m \kappa_t k^2 x_{ES}^6} = 4.4 \times 10^6 \, s \times \frac{a^3 m_{2.8}}{\alpha_{-10} \kappa_t k^2 (x_{ES}/0.5)^6} \qquad (18)$$

For the low-density medium around the spinar, if $R_t = R_{light}$, the level of the magneto-dipole losses is less and the transient object lives longer:

$$L_{md} = \frac{2\alpha_m \kappa_t k^8 x_{ES}^{18}}{a^8} \frac{c^5}{G} = 2.8 \times 10^{44} \, \text{erg/s} \times \frac{\alpha_{-10} \kappa_t k^8 (x_{ES}/0.5)^{18}}{a^8} \qquad (19)$$

$$t_{plato} = \frac{1}{2} \frac{Rg}{c} \frac{a^6}{k^5 x_{ES}^{12} \kappa_t \alpha_m} = 2.8 \times 10^8 \, s \times \frac{a^6 m_{2.8}}{\alpha_{-10} \kappa_t k^5 (x_{ES}/0.5)^{12}}. \qquad (20)$$

The following parameters have been introduced. The moment of the inertia of the spinar is expressed using the dimensionless parameter $k$ as $I_{sp} = k M_{sp} R_{sp}^2$. Parameter $x_{ES}$ is an implicit function of the equation of state and is calculated as $x_{ES} = R_{sp}/R_{cor}$ during the plateau stage. If the intermediate object has a very high mass and is mainly supported against gravity by its rotation, then $x_{ES} = 1$. In the context of a merger of two neutron stars with masses of ~ 1.4 $M_\odot$, we obtain $x_{ES}$ in the range 0.5-0.6, if the dimensionless Kerr parameter of the transient object is around 1.5.

Parameter $\alpha_m$ describes the strength of the magnetic field and is the ratio of the magnetic energy of the spinar to its gravitational energy:

$$\alpha_m \equiv \frac{U_m}{GM_{sp}^2/R_{sp}}, \qquad (21)$$

where $M_{sp}$ and $R_{sp}$ are the spinar's mass and radius, $U_m = B^2 R^3 /6$ is the energy of the magnetic field, and the magneto-dipole momentum $\mu = (BR^3)/2$. During the course of the collapse, the magnetic flux through the collapsar and, hence, parameter $\alpha_m$ remain constant. The strength of the magnetic field can be written as

$$B = 1.6 \times 10^{12} G \times \frac{\alpha_{-10}^{1/2} m_{2.8}}{(R_{sp}/50 \, \text{km})} \qquad (22)$$

where $\alpha_{-10} = \alpha_m/10^{-10}$ and $m_{2.8} = M_{sp} / (2.8\, M_\odot)$

In the case of the event GW170817 / GRB170817A, we most likely see the off-axis (lateral) gamma radiation, which can be thousand times weaker. Indeed, GRB170817A consists of the initial short (~ 0.5 sec), hard (although not as hard as a usual short GRB) event, after which the softer tail was observed for several seconds.

The total energy reported is $E_{iso,total} = (5.35\pm1.26)\, 10^{46}$ erg/s, and the peak luminosity $L_{iso,peak} = (1.62\pm0.57)10^{47}$ erg/s (LVC, GBM, INTEGRAL, 2017 in prep).

One should take into account that the expanding matter overcasts the central object from an observer. Notably, the matter of the shell expands with a relativistic velocity (Cowperthwaite et al. 2017, Malesani et al. 2017, Lipunov 2017f): $v \sim 0.3\, c$.

Let us find the time when the shell becomes transparent to the X-ray radiation. The optical thickness in the hard X-ray range is determined by the Thompson scattering:

$$\tau = \kappa_T\, \rho\, \Delta R = 1, \qquad \rho \sim M_{sh}/4\pi R^2\, \Delta R\,, \quad R = V_{sh}\, t = (1/3)\, ct \qquad (23)$$

where $M_{sh}$ is the shell's mass, $\kappa_T = 0.36$ cm$^2$/g is the cross-section of the Thomson scattering, $\Delta R$ and $V_{sh}$ are the thickness and velocity of the shell. We obtain that

$$t_{clear} \sim 10^5\, s\, \sqrt{M_{sh}/0.01 M_{sun}}\,. \qquad (24)$$

In connection with the last formula, it is of great interest that a point X-ray source was detected by Chandra 9 days after the event in the direction of the merger site (Troja et al. 2017a, Troja et al. 2017b) but it was not found earlier (Margutti et al. LVC GCN 21648). Assuming the Chandra sensitivity of its Advanced CCD Imaging Spectrometer (4 x $10^{-15}$ ergs/cm$^2$/s for the exposure $10^4$ s in 0.4-6.0 keV, http://cxc.cfa.harvard.edu/cdo/about_chandra/), we estimate the lower limit on the flux for the time of the total exposure of 50 ks as 9 x $10^{-15}$ erg/cm$^2$/s/. 15 days after the merger event, the point X-ray source was detected with the flux $4.5\, 10^{-15}$ erg/cm$^2$/s.

In Fig. 7 we plot the reported X-ray observations. A spinar model producing a neutron star can fit the observed data. The magnetic field has to be $5\times10^{14}$ G at the plateau stage and about $10^{15}$ G during the magneto-dipole decay. A factor $\eta = 3\cdot10^{-4}$. of conversion from the magneto-dipole luminosity to the observed X-rays has to be involved.

There is one more possibility – spinar, - as a disc of superfluid accretion, see Appendix

## 9. Conclusion.

This review article is written in the footsteps of the historical discovery of the merger of neutron stars in the gravitational, gamma, X-ray, infrared, and optical ranges on August 17, 2017. Thanks to the detection in the optical range, it is possible to establish the exact coordinates of the event occurred in the NGC 4993 galaxy at a distance of ~ 40 Mpc from the Earth. The very fact of the discovery of such a close merger is in the excellent agreement with the results of the population synthesis, based not only on the use of "new" trends, but also on the numerous observational examples of different manifestations of neutron stars, from X-ray pulsars to double radio pulsars, and not yet discovered black holes in pairs with pulsars, the absence of which also plays an important role in determining the hidden parameters of the evolution of binary stars.

We note that in this article we consider only those neutron stars that are born as field stars of galaxies and are not re-cycled pulsars (see Özel and Freire 2016). A number of authors, especially in recent years, insist that the binary stars in globular clusters could play an important role in the merger rate of neutron stars (Belczynski et al., 2002). We will only note that the evolution of such binary systems, starting with the issue with their formation (typically, by random captures) is poorly understood at the present time to draw some definite conclusions even with an accuracy of factor 2-3, the accuracy achieved in the population synthesis of field stars, which is incomparably richer in the sense of observed diversity of the evolutionary stages of the binary stars (Grishchuk et al., 2001).

We have shown that, at the star formation rate of NGC 4993 galaxy, which is estimated by the track is constructed as an example for such binary.

The general characteristics of the optical flare, discovered independently by several optical projects, Swope and MASTER (Lipunov et al. 2017), agree with the previously predicted properties of the kilonova phenomenon (Li and Paczynski 1998, Metzger et al. 2010; Tanvir et al. 2013, Berger et al. 2013). We took a step further and found that two more candidates for kilonovae: the observed earlier GRB 130603B (Tanvir et al. 2013) and GRB 080503 (Perley et al. 2009) have approximately the same luminosity. For example, the luminosity in the kilonova of GW170817 differs from the luminosity of GRB 080503, located at a distance of 2 orders of magnitude farther, that is, 4 orders of magnitude less in flux, by only 2 times! Note that the luminosity or the absolute stellar magnitude of a kilonova is directly related to the mass of the nuclear matter discarded during the merger. The chemical composition of all neutron stars is the same: `they do not have a chemical composition'. The discarded mass is determined only by the initial masses of the merging stars, which, judging by the observations of the double radio pulsars, differ only by a couple of tens of percent (Özel and Freire 2016).

If the hypothesis of a kilonova as the standard candle is confirmed in the future, we will get an independent method for estimating the distance to the merger site. This can be important because the binary neutron stars can leave the galaxies and during the Hubble time can move away by megaparsecs. At the same time, the spectra of kilonovae contain a huge number of lines of heavy radioactive elements, which are not so easy to identify (Shara et al.2017, Drout et al.2017).

The most important question is the nature of the remnant after the neutron stars merger: a black hole or a neutron star? In both scenarios, a temporary object – a spinar or an accretion disk - may appear. We note here that the accretion disk around the newly formed black hole can be superfluid and, thus, can exist for a long time of the order of one month. A good observational test can be made by an observation in the X-ray range, after the discarded envelope becomes transparent with respect to the Thompson scattering. A detailed study of these issues continues and, perhaps, the final clarity in them will be provided sooner by new observations of the merging of neutron stars.

**Anknowlegment**

**This paper** is supported in part by the Development Programm of Lomonosov Moscow State University, Moscow Union OPTICA, Russian Science Foundation 16-12-00085, Russian Foundation of Fundamental Research 15-02-07875, and National Research Foundation of South Africa. We are especially grateful to S.M.Bodrov for his long years MASTER's support.

APENDIX A.

To explain the duration of the activity of the object formed after the coalescence of two neutron stars, yet another scenario can be schematically proposed.

It cannot be ruled out that the matter of two merged neutron stars forms a differentially-rotating disk-like object. Let us consider it's viscous dissipation. It is known, that, as the matter in a neutron star cools down below T ~ $10^9$ K, neutrons and protons in its core dwell in the superfluid state (e.g., Shapiro and Teukolsky 1983).

Generally, the viscosity inside the neutron star is the combination of interactions between protons, neutrons, muons, and electrons and depends on whether or not the nucleons are in the superfluid state (e.g. Sternin and Yakovlev 2008; Kolomeitsev and Voskresensky 2015). When the neutrons become superfluid, they cease to contribute to the dissipation. The protons are also expected to make the transition to a superfluid state at about the same temperatures and densities. Consequently, only the normal electrons are able to transport the momentum and energy by scattering, and only electron-electron scattering will occur. As noticed by Cutler and Lindblom (1987), contrary to the experience with other superfluids like He4, neutron star matter becomes more viscous in the superfluid state than it was in the normal state.

Leptons in the neutron star matter are not superfluid. The lepton shear is dominated by the electron term. The lepton shear viscoity was theoretically calculated and approximated by analytic formulas by Flowers and Itoh (1979), Cutler and Lindblom (1987); Sternin and Yakovlev (2008); Kolomeitsev and Voskresensky (2015); Gusakov et al (2014). We adopt the following approximation by Gusakov et al (2014) for the dynamic coefficient of viscosity that uses results of Sternin and Yakovlev (2008):

$$\eta_e = 6 \cdot 10^{20} \quad \rho_{15}^2 \quad T_8^{-2} \quad g/cm/s, \qquad (25)$$

Where $T_8 = T/10^8$ K, $\rho_{15} = \rho/10^{15} g/cm^3$.

Note, that an uncertainty factor of several is quite possible in the formula above (Gusakov et al 2014).

The viscous time scale in the disk-like object can the estimated as

$$t_{vis} = R^2 \rho / \eta_e = 1.5 \cdot 10^7 s \times R_{30}^2 \quad T_8^2 \quad \rho_{15}^{-1} \qquad (26)$$

The corresponding energy release from the matter falling into the central gravitational well during the viscous evolution is

$$L \sim \frac{G M^2}{R \, t_{vis}} \sim 7 \cdot 10^{46} erg/s \times R_{30}^{-1} \, t_7^{-1}, \qquad (27)$$

Where $R_{30} = R/30 km$ and $t_7 = t/10^7 s$ and $M = 2.8 M_\odot$.